\documentclass[twocolumn,eqsecnum,aps,showpacs]{revtex4}
\usepackage{graphics}
\newcommand{\be}{\begin{equation}}
\newcommand{\ee}{\end{equation}}
\newcommand{\bea}{\begin{eqnarray}}
\newcommand{\eea}{\end{eqnarray}}
\begin{document} 
\begin{flushleft}
{\em This article has been submitted to \\ the Journal of Chemical Physics.}
\end{flushleft}
\title{Nucleation of stable cylinders from a metastable lamellar phase \\ 
	in a diblock copolymer melt}
\author{Robert A. Wickham and An-Chang Shi} 
\affiliation{Department of Physics and Astronomy, 
McMaster University, 
Hamilton, Ontario L8S 4M1, Canada}
\author{Zhen-Gang Wang}
\affiliation{Division of Chemistry and Chemical Engineering, 
California 
Institute of Technology, Pasadena, California 91125}
\date{\today}
%
%
\begin{abstract}
The nucleation of a droplet of stable cylinder phase from a metastable
lamellar phase is examined within the single-mode approximation to the
Brazovskii  model   for  diblock  copolymer  melts.   By  employing  a
variational ansatz  for the  droplet interfacial profile,  an analytic
expression  for  the  interfacial   free-energy  of  an  interface  of
arbitrary  orientation between  cylinders and  lamellae is  found. The
interfacial free-energy is anisotropic, and is lower when the cylinder
axis is  perpendicular to  the interface than  when the  cylinders lie
along the  interface.  Consequently,  the droplet shape  computed {\em
via} the  Wulff construction is  lens-like, being flattened  along the
axis  of the  cylinders.  The  size of  the critical  droplet  and the
nucleation barrier are  determined within classical nucleation theory.
Near the lamellar/cylinder  phase boundary, where classical nucleation
theory  is applicable,  critical  droplets of  size 30--400  cylinders
across with aspect  ratios of 4--10 and nucleation  barriers of 30--40
$k_B T$ are typically found.   The general trend is to larger critical
droplets, higher aspect ratios  and smaller nucleation barriers as the
mean-field critical point is approached.
\end{abstract}
\maketitle
%
%
\section{Introduction}

The  decay of  a metastable  state {\em  via}  the thermally-activated
formation and  subsequent growth of droplets of  the equilibrium phase
is known as nucleation. If  the nucleation rate is small the formation
of critical droplets  --- those droplets large enough  to overcome the
nucleation  barrier   and  grow  ---   can  be  considered  to   be  a
quasi-equilibrium  process.  In  the classical  homogeneous nucleation
scenario,  the  critical  droplet  size  and  nucleation  barrier  are
determined  from a balance  between the  droplet interfacial  and bulk
free-energies \cite{NUCLEATION}.

Nucleation  from a  disordered  initial state  to  a disordered  final
state,   as   occurs   in   the  liquid-gas   transition,   has   been
well-studied. The  study of  nucleation when ordered,  periodic phases
are  present  is challenging,  and  has  been  mainly studied  in  the
materials  science   context  of  crystal   growth  \cite{OFFERMAN02}.
However there is also a  large class of soft-materials where competing
interactions select  a microscopic  length-scale and lead  to ordered,
periodic  microphases  \cite{SEUL95}.  Examples  include  ferrofluids,
thin-film     magnetic    garnets,    Rayleigh-B$\acute{\mbox{e}}$nard
convection,  Langmuir  monolayers and  block  copolymers. Whether  the
material is  hard or  soft, the challenge  to modelling  nucleation in
systems with microstructure arises mainly from the need to incorporate
the  various length-scales  into  the  theory. In  the  case of  block
copolymer melts these length-scales  are the interfacial width between
microdomains, the period of the microdomains, the width of the droplet
interface,  and the  critical droplet  size.  Another  complication is
that  the  symmetry of  the  underlying  microstructure  will lead  to
anisotropic droplet interfacial free-energies and anisotropic critical
droplet  shapes.   Furthermore,   the  droplet  interfacial  structure
becomes complicated as the different microdomain symmetries merge into
each other.

In  this  paper  we  will  focus on  diblock  copolymer  melts,  whose
equilibrium       phase       behaviour       is       well-understood
\cite{LEIBLER80,MATSEN94,MATSEN02}.   At high temperatures  the system
is disordered,  while at  lower temperatures several  ordered periodic
phases ---  lamellar, hexagonally-packed cylindrical, body-centred-cubic
spherical, and  gyroid --- compete for stability.  This understanding of
the equilibrium properties of diblock copolymer melts provides a solid
foundation  on  which  to  study  the kinetics.   Within  the  diblock
copolymer phase diagram we  will study nucleation near the first-order
lamellar/cylinder phase boundary following a temperature jump into the
cylinder  phase.  This  is  the  simplest  order-order  transition  to
examine, yet  it contains all  of the challenges and  issues mentioned
above.  The goal of this work is  to compute the size and shape of the
critical  droplet and  the  free-energy barrier  to  nucleation for  a
stable cylinder phase nucleating from a metastable lamellar background
in a diblock copolymer melt.

Previous theories for the kinetics of order-order transitions in block
copolymers have  examined the behaviour near  the order-order spinodal
where, in contrast to nucleation, a particular mode of fluctuation out
of  the metastable state  becomes unstable,  and the  metastable state
transforms uniformly  throughout the entire sample into  a more stable
state   \cite{CAHN59,   LARADJI97,   SHI99,  QI97,   QI98,   MATSEN98,
MATSEN01}.  The  recent   successful  application  of  self-consistent
field-theory to block  copolymers relies on efficient reciprocal-space
methods    for    treating    ordered,   uniform    periodic    phases
\cite{MATSEN94,LARADJI97,  SHI99}. In  nucleation, this  uniformity is
absent,  and  this has  limited  the  application of  reciprocal-space
methods to  the problem.  The  kinetics of various  order-disorder and
order-order  transitions  have  been  studied  numerically  in  Refs.\
\cite{BAHIANA90, ROLAND90,  QI96, BOYER02, FRAAIJE93}.  The relatively
small three-dimensional system  sizes currently accessible numerically
make it  impossible to  handle the various  length scales  involved in
nucleation.  Thus  direct numerical  studies are generally  limited to
examining transitions  across the entire  system at once  ({\em i.e.}\
the spinodal regime).  It is not  clear what role, if any, the kinetic
pathways  and uniform,  periodic  intermediate states  found in  these
studies  play in  the nucleation  of compact  droplets of  one ordered
phase in another.

The  existing  theoretical work  focusing  on  nucleation in  polymeric
systems considers different systems and situations than
the one considered here. Wood and Wang recently examined nucleation in
polymer  blends   using  self-consistent  field-theory  \cite{WOOD02}.
Fredrickson  and Binder \cite{FREDRICKSON89}  and later  Hohenberg and
Swift \cite{HOHENBERG95}  examined the nucleation of  a lamellar phase
from  an initially  disordered  phase in  melts  of symmetric  diblock
copolymers.  While   Ref.\  \cite{FREDRICKSON89}  only   considered  a
spherical  droplet, Ref.\ \cite{HOHENBERG95}  found that  the critical
droplet shape  should be anisotropic,  with the lamellar  normal along
the long-axis of the droplet. Two-dimensional numerical simulations by
Nonomura and Ohta  have recently examined the dynamics  of droplets of
lamellae    nucleating     in    a    metastable     cylinder    phase
\cite{NONOMURA01}. This is close to the situation considered here, and
we will discuss it in more detail later.

Early experimental work by Hajduk  {\em et al.}\ observed a reversible
thermotropic transition  between the lamellar and cylinder  phase in a
polystyrene-poly(ethene-{\em     co}-butene)     diblock     copolymer
\cite{HAJDUK94}.   The transition pathway  from lamellae  to cylinders
indicated  by their data  suggests that  they were  observing spinodal
decomposition rather than nucleation.   Sakurai {\em et al.}\ examined
the  reverse  transition  in  poly(styrene-{\em  block}-butadiene-{\em
block}-styrene) triblock copolymers,  where a non-equilibrium cylinder
phase produced by a selective  solvent transforms into a stable lamellar
phase upon annealing \cite{SAKURAI93}.  It appears that they were also
observing spinodal  decomposition, although the  existence of isolated
regions  of   the  metastable  cylinder  phase  in   their  system  is
interesting.   Floudas {\em  et  al.}\ have  studied  kinetics of  the
lamellar-to-cylinder transition  in poly(isoprene-{\em block}-ethylene
oxide) where the  ethylene oxide block is crystalline  in the lamellar
phase \cite{FLOUDAS00}.  This  crystallinity, the observed rapidity of
the transition and  the lack of specific information  about the nuclei
limit  comparison  with  the  present work.   Experimental  work  that
directly addresses nucleation is  limited to studies of order-disorder
transitions  between either  the  disordered and  the lamellar  phases
\cite{HASHIMOTO95} or, as will be  discussed in more detail later, the
disordered and the cylinder phases \cite{DAI96,BALSARA98a,BALSARA98b}.
These experiments  suggest that there is an  incubation time, followed
by nucleation  of anisotropic droplets  of the ordered phase  from the
disordered phase.

To  study the  decay  of a  metastable  lamellar state  {\em via}  the
nucleation  of  compact  droplets  of  cylinder phase  we  employ  the
Landau-Brazovskii  model \cite{SHI99,BRAZOVSKII75,KATS93}.  This model
is   appropriate  to   systems   with  a   weak   modulation  of   the
order-parameter, or  weak segregation, as occurs  for block copolymers
near the mean-field critical point. Other phenomena studied using this
model   include  weak  crystallization,   the  nematic   to  smectic-C
transition      in     liquid      crystals      \cite{KATS93}     and
Rayleigh-B$\acute{\mbox{e}}$nard    convection   \cite{SWIFT77}.   The
Landau-Brazovskii  free-energy  can  be  derived from  the  many-chain
Edwards  Hamiltonian for  diblock copolymers  in the  weak segregation
limit  \cite{LEIBLER80, OHTA86}.  This  important  connection allows  us 
to  move
beyond phenomenology in this study and make specific predictions about
the  size and  shape  of  the critical  droplets,  and the  nucleation
barrier.

We apply the single-mode  approximation, accurate at weak segregation,
to the  Landau-Brazovskii model, which  results in an  amplitude model
studied         previously        in         different        contexts
\cite{NONOMURA01,TURNER94,PODNEKS96,GOVEAS97}.   Since  we  will  work
within the framework of classical homogeneous nucleation theory, a key
ingredient in  our theory is the droplet  interfacial free-energy.  We
compute the interfacial free-energy  for a planar interface separating
coexisting lamellar and cylinder  phases from the amplitude model.  By
using a  variational approach, we obtain an  analytical expression for
the interfacial  free-energy for interfaces  of arbitrary orientation.
Previous  studies  of interfacial  free-energy,  which were  primarily
numerical, have  examined only selected interfacial  geometries in two
dimensions  \cite{NETZ97}.   With  our  results  for  the  interfacial
free-energy,  we  can  compute   the  droplet  shape  from  the  Wulff
construction \cite{WULFF01}.  The Wulff  construction has been used to
study   the    anisotropic   droplet   shape    arising   during   the
disorder-to-lamellar transition  in Ref.\ \cite{HOHENBERG95}  and more
recently in Ref.\ \cite{BALSARA02}. We  then apply our results for the
interfacial free-energies and droplet  shape to calculate the critical
droplet  size  and   nucleation  barrier  from  classical  homogeneous
nucleation  theory. 
%
%
\section{Theory}
\subsection{Landau-Brazovskii Theory}
\label{SEC:LB}

We consider an  incompressible melt of $n$ AB  diblock copolymers in a
volume $V_0$ at a temperature $T$.  The total degree of polymerization
of the  diblock copolymer 
is $N$. The  monomer density is  thus $\rho_0 = n  N /
V_0$.  The degree  of polymerization of the A block  is $f_A N$, where
$0 \leq f_A \leq 1$.   We employ the Landau-Brazovskii theory for weak
crystallization    to     study    nucleation    in     this    system
\cite{BRAZOVSKII75,KATS93,SHI99}.  The position-dependent 
order-parameter, $\phi ({\bf r}_0)$, 
in this theory  is defined as 
the deviation of the  normalized A monomer concentration,
$\phi_A$, from the uniform state, and  is given by $\phi ({\bf r}_0) =
\phi_A ({\bf  r}_0)- f_A$. The Landau-Brazovskii  free-energy, $F_0$, is
an expansion in terms of this order-parameter,
%
%
\bea
f_0 \equiv \frac{F_0}{ n k_B T} & = & \frac{1}{V_0}
\int d {\bf r}_0 \left\{ \frac{\xi_0^2}{8 q_0^2} [ ( \nabla^2 + q_0^2) \phi ]^2
+ \frac{\tau_0}{2} \phi^2 \right.
\nonumber \\ 
& & \left. \hspace{.5 in}
- \frac{\gamma_0}{3!} \phi^3 
+ \frac{\lambda_0}{4!} \phi^4 \right\}.
\label{EQ:LB0}
\eea
In Eq.\ (\ref{EQ:LB0})  $f_0$ is the Landau-Brazovskii free-energy
per  block copolymer in  units of  $k_B T$,  $\xi_0$ is  the  bare correlation
length,  $q_0$ is  the critical  wavevector, $\tau_0$  is  the reduced
temperature,   and   $\gamma_0$    and   $\lambda_0$   are   expansion
coefficients.  For stability, $\lambda_0  > 0$.  The Landau-Brazovskii
free-energy   is   able   to   account  for   the   observed   diblock
copolymer microstructures \cite{SHI99}.

The Landau-Brazovskii  free-energy can be derived  from the many-chain
Edwards Hamiltonian for diblock  copolymers following the method of 
Leibler \cite{LEIBLER80} and Ohta
and Kawasaki  \cite{OHTA86}.  Like many authors, we  have assumed that
the  third- and  fourth-order expansion  coefficients are  local.  The
truncation of  the order-parameter expansion  at fourth-order  in Eq.\
(\ref{EQ:LB0})  and the  assumption of  a single  dominant wavevector,
$q_0$,  in  the gradient  term  are  valid if  the  system  is in  the
weak-segregation  limit  near  the  disorder-to-order  spinodal  and  the
mean-field 
critical  point.  Since, in  mean-field theory,  the lamellar-cylinder
transition occurs in this region, the Landau-Brazovskii free-energy is
appropriate to our study.

It is convenient to rescale Eq.\ (\ref{EQ:LB0}) by expressing lengths in 
units of 
$q_0^{-1}$ and the free-energy is in units of $\lambda_{0}$. Under the 
rescalings
%
%
\bea
{\bf r } & = & q_0 {\bf r_0} \\
V & = & q_0^3 V_0 \\ 
f & = & \frac{f_0}{\lambda_0} \\ 
\xi^2 & = & \frac{(q_0 \xi_0)^2}{4 \lambda_0} \\
\tau & = & \frac{\tau_0}{\lambda_{0}} \\ 
\gamma & = & \frac{\gamma_{0}}{\lambda_{0}}
\eea
the Landau-Brazovskii free-energy becomes
%
%
\be
f = \frac{1}{V}
\int d {\bf r} \left\{ \frac{\xi^2}{2} [ ( \nabla^2 + 1) \phi ]^2
+ \frac{\tau}{2} \phi^2 - \frac{\gamma}{3!} \phi^3 
+ \frac{1}{4!} \phi^4 \right\}.
\label{EQ:BL}
\ee

One of the goals of this work is to calculate the
critical size, shape,  and nucleation barrier for droplets of stable cylinder
phase nucleating from a metastable  lamellar phase in terms of 
experimentally measurable quantities. To do this we must
relate  the  parameters  $q_0$,  $\xi_0$,  $\tau_0$,  $\gamma_0$,  and
$\lambda_0$  appearing   in  Eq.\  (\ref{EQ:LB0}) to  such quantities.
   By  following   the   derivation  of   Eq.\
(\ref{EQ:LB0}) in  Ref.\ \cite{OHTA86}, we  can express these  parameters in
terms of  $\chi$, $N$, $f_A$ and  $R_g$ that appear  in the many-chain
Edwards model.  Here $\chi$ is the Flory-Huggins interaction parameter
characterizing  the repulsion between  A and  B monomers.  The diblock
copolymer radius of gyration, $R_g$, is defined through
%
%
\be
R_g^2 = \frac{N b^2}{6},
\ee
where $b$ is the Kuhn statistical segment length. We have 
\bea
q_0^2 & = & \frac{x^*}{R_g^2} 
\label{EQ:Q0} \\
\xi_0^2 & = & 4 x^* c R_g^2 \\
\tau_0 & = & 2 [ (\chi N)_s - \chi N ] \\
\gamma_0 & = & - N \Gamma_3 \\
\lambda_0 & = & N \Gamma_4 (0,0).
\eea
The notation follows that  of Leibler \cite{LEIBLER80}. In particular,
$x^*$  is the  position  of  the minimum  of  the function  $F(x,f_A)$
appearing in  the scattering  function of Ref.\  \cite{LEIBLER80}. The
disorder-to-order  spinodal,  $(\chi  N)_s$,  and the  quantity,  $c$,  are
defined through
\bea
(\chi N)_s & = & \frac{1}{2} F(x^*,f_A) \\
c & =  & \frac{1}{2} \left. \frac{d^2 F(x,f_A)}{d x^2} \right|_{x = x^*}.
\eea
Finally, the  vertex functions $\Gamma_3$ and  $\Gamma_4 (0,0)$, which
are  functions  of  $f_A$,  are computed  in  Ref.\  \cite{LEIBLER80}.
In Table \ref{TAB:PARA} 
we  reproduce the values  tabulated by Fredrickson  and Helfand
for  these  mean-field  parameters  in  the
weak-segregation region \cite{FREDRICKSON87}. We will take $f_A \leq 0.5$ in
this paper without loss of generality.
%
%
\begin{table}[h]
\caption{Mean-field model parameters in the weak-segregation region, from Ref. 
\cite{FREDRICKSON87}.} 
\label{TAB:PARA}
\vspace{.2 in}
\begin{tabular}{|llllll|}
\hline 
$f_A$ & $x^*$ & $(\chi N)_s$ & $c$ &  $N \Gamma_3$ & $N \Gamma_4 (0,0)$ \\
\hline 
0.50 \hspace{.2 in} & 3.7852 \hspace{.2 in} & 10.495  \hspace{.2 in} & 0.4812 
\hspace{.2 in} & 0.0 	& 156.56 \hspace{.2 in} \\
0.45 & 3.7995  & 10.698 & 0.4844 & -8.608 & 169.19 \\
0.40 & 3.8433  &  11.344 & 0.4945 & -18.81 \hspace{.2 in} & 212.22 \\
\hline
\end{tabular}
\end{table}

In terms of these parameters, the scaled quantities appearing in 
Eq.\ (\ref{EQ:BL}) are
\bea
\xi^2 & = & \frac{(x^*)^2 c}{N \Gamma_4 (0,0)} \\
\tau & = & \frac{2 [(\chi N)_s - \chi N]}{N \Gamma_4 (0,0)} 
\label{EQ:TAUCHI} \\
\gamma & = & - \frac{N \Gamma_3}{N \Gamma_4 (0,0)}.
\label{EQ:GAMVERT}
\eea
For example, for $f_A=0.5$ we have  $\xi^2 = 0.0440$ and $\gamma = 0$,
while  for $f_A  =  0.45$ we  have  $\xi^2 =  0.0413$  and $\gamma  =
0.0509$.  When $\chi N  = 11.0$,  $\tau =  -6.45 \times  10^{-3}$ for
$f_A=0.5$ and  $\tau =  -3.56 \times 10^{-3}$  for $f_A =  0.45$. The
free-energies  of  the   equilibrium  lamellar  and  cylinder  phases
computed in the weak-segregation  regime using these parameter values
differ by  only a  few percent from  state-of-the-art self-consistent
field-theory calculations \cite{SCFT}.
%
%
\subsection{Geometry of the Droplet and the Single Mode Approximation}
\label{SEC:SMODE}

In  principle, to describe  a critical  droplet of  stable cylindrical
phase in a background of metastable lamellar phase one needs to find a
saddle-point solution to the Euler-Lagrange equations corresponding to
Eq.\   (\ref{EQ:BL}).   In  practice,   for  the   three-dimensional,
anisotropic droplets we consider, finding  such a solution by a direct
numerical method is not, at  present, feasible due to the large system
sizes required to resolve the  various length-scales in the problem ---
the  A/B interfacial  width,  the microstructure  period, the  droplet
interfacial width  and the  droplet size. Thus,  to make  progress, we
need to make some basic  assumptions about the geometry of the droplet
and the form of the order-parameter.

\begin{figure}
\resizebox{2.43in}{1.82in}{\includegraphics{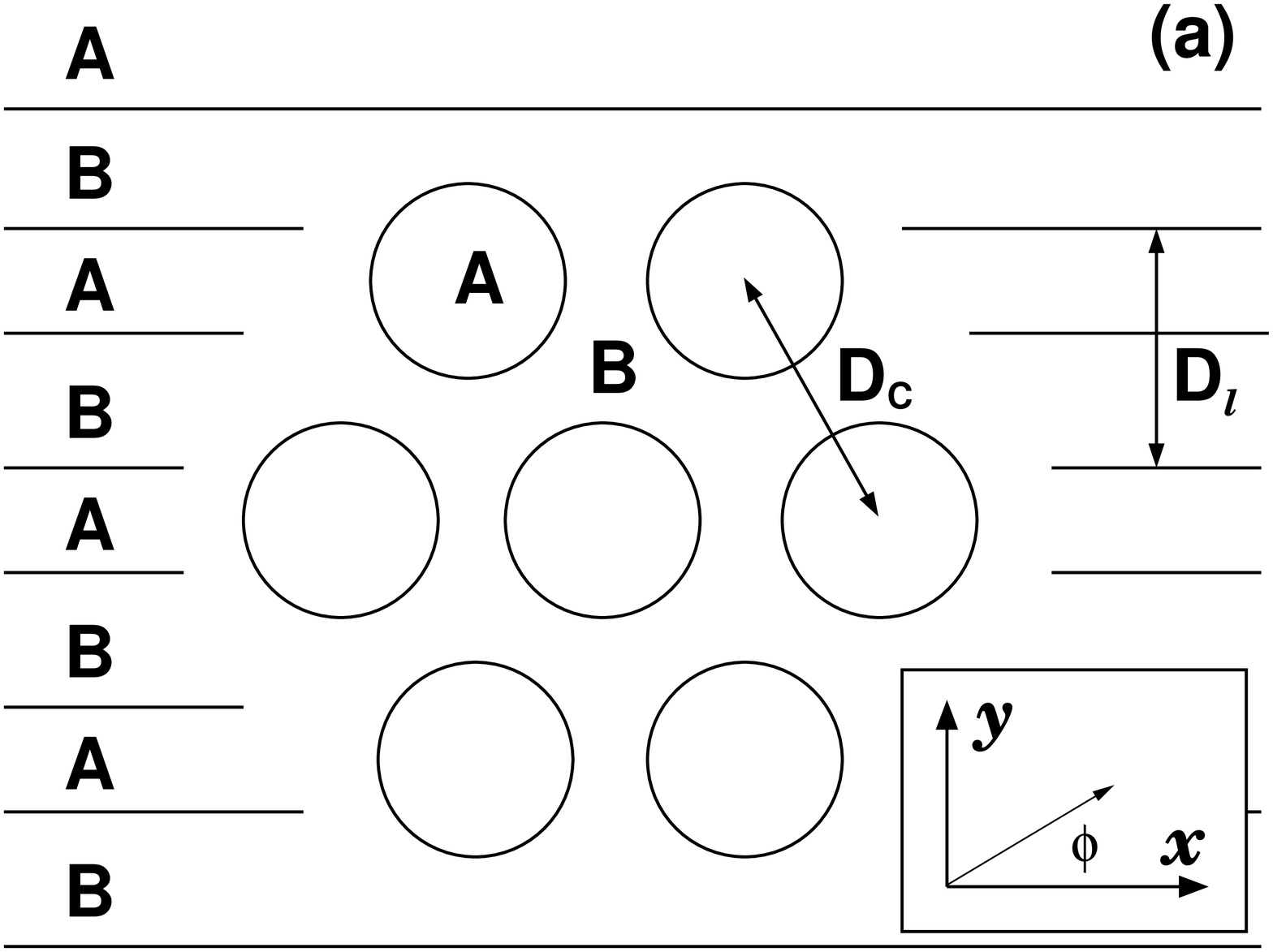} }
\\
\vspace{0.5 in}
\resizebox{2.43in}{1.82in}{\includegraphics{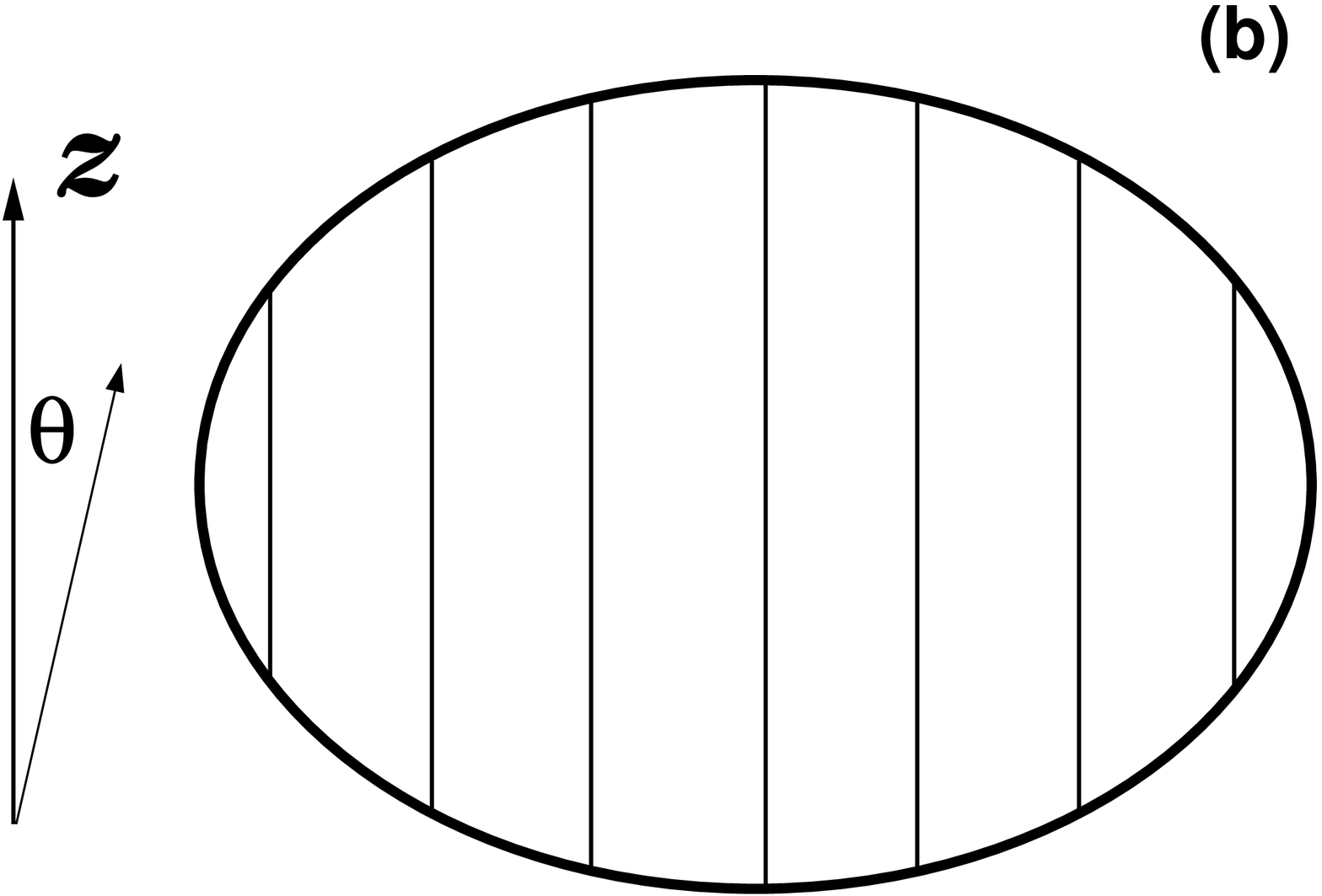} }
\caption{ 
Schematic  picture of the  droplet geometry. In  reality the
critical droplet is much larger  than shown.  (a) View of the $x$--$y$
plane. The  lamellar normal  is along the  $y$-axis and  the cylinders
form a hexagonal lattice inside  the droplet with their axes along $z$
(out of the page). The  lamellar period is $D_l$, the distance between
cylinders  is $D_c$. The  azimuthal angle,  $\phi$, is  indicated. (b)
Perpendicular view  to (a), showing  the orientation of  the cylinders
relative  to the  droplet  interface. The  polar  angle, $\theta$,  is
indicated. }
\label{FIG:GEOMETRY}
\end{figure}

We consider a  droplet of stable cylindrical phase  in a background of
metastable   lamellar  phase   with  the   geometry  shown   in  Fig.\
\ref{FIG:GEOMETRY}.   Inside the  droplet, the  cylinders  are ordered
into a hexagonal  lattice and aligned along the  $z$-axis. Outside the
droplet, the lamellar normal is directed along the $y$-axis. We assume
an epitaxial relation between the cylinders and the lamellae. Matsen's
work on the cylinder/gyroid and cylinder/sphere transitions shows that
these order-order  transitions proceed approximately  epitaxially, and
that imposing an epitaxial relation  results in only a minor change in
the  energetics  of   the  transition  \cite{MATSEN98,MATSEN01}.  This
suggests that  the epitaxial  assumption may also  have only  a slight
effect     on    the     energetics    of     the    lamellar/cylinder
transition.   Furthermore,   experiments    indicate   that   at   the
lamellar/cylinder transition  the cylinders will form  with their axis
in    the   plane    of    the   layers,    as    we   have    assumed
\cite{SAKURAI93,HAJDUK94}. The lamellar period, $D_{l}$, is related to
$q_0$ by
%
%
\be
q_0 = \frac{2 \pi}{D_l}.
\ee
The  distance between  cylinders, $D_{c}$,  is epitaxially  related to
$D_l$  by  $D_c  = 2 D_l/  \sqrt{3}$.
%
%

The orientational epitaxy enables  us to describe both the cylindrical
and the lamellar order in  terms of a single set of reciprocal-lattice
vectors ${\bf G}_i$, indexed by the  integer $i$. Since we work in the
weak-segregation regime it is  sufficient to restrict ourselves to the
first  mode, those vectors  with $|{\bf  G}_i|=q_0$ (in  scaled units,
$|{\bf G}_i|=1$), instead  of using the complete set  of vectors ${\bf
G}_i$.   In  the  single-mode  approximation, the  order-parameter  is
written as
%
%
%
\be
\phi({\bf r})  = 2 a_1 ({\bf r}) \mbox{ } \cos( {\bf G}_1 \cdot {\bf r})
+ 2 a_2 ({\bf r}) \mbox{ } [ \cos( {\bf G}_2 \cdot {\bf r}) + 
\cos( {\bf G}_3 \cdot {\bf r}) ], 
\label{EQ:OPDECOMP} 
\ee
where $a_1$ and $a_2$ are  spatially-dependent amplitude functions
which define the droplet geometry and the ${\bf G}_i$ are given (in scaled
units) by
%
%
\bea 
{\bf G}_1 & = & {\bf \hat{y}} \\ 
{\bf G}_2 & = &  \frac{1}{2} \left( - \sqrt{3} \mbox{ }  
{\bf \hat{x}} + {\bf \hat{y}} \right) \\ 
{\bf G}_3 &  = & \frac{1}{2} \left( -
\sqrt{3} \mbox{ } {\bf \hat{x}}  - {\bf \hat{y}} \right).  
\eea
Pure lamellar  order is described by  $a_1 \neq 0$ and $a_2 =  0$, 
pure cylindrical order by $a_1=a_2 \neq 0$.

If we assume the amplitudes $a_1$  and $a_2$ are slowly varying on the
scale of $D_c$,  we can separate the  length-scale for variations
in the amplitude, the droplet interfacial width, from the length-scale
of the underlying microstructure. Within this slowly-varying amplitude
approximation, the  Landau-Brazovskii free-energy, Eq.\ (\ref{EQ:BL}),
can be written in terms of the amplitudes as
%
%
\bea
f & = & \frac{1}{V} \int d{\bf r} \Big\{  \xi^2 (\nabla^2 a_1)^2 +
2 \xi^2 (\nabla^2 a_2)^2 
+ 4 \xi^2 [({\bf G}_1 \cdot \nabla a_1)^2 
\nonumber \\ & &
+ ({\bf G}_2 \cdot \nabla a_2)^2
+ ({\bf G}_3 \cdot \nabla a_2)^2 ] 
+ \tau (a_1^2 + 2a_2^2) 
\nonumber \\ & &
- 2 \gamma a_1 a_2^2 
+ \frac{1}{4}(a_1^4 + 6 a_2^4 + 8 a_1^2 a_2^2) \Big\}.
\label{EQ:LBSINGLE}
\eea
Extremization   of  this   free-energy  produces   two  Euler-Lagrange
equations for the amplitudes,
%
%
\bea
2 \xi^2 \nabla^4 a_1  
- 8 \xi^2 ({\bf G}_1 \cdot \nabla)^2 a_1 + v_1 (a_1, a_2) &  = & 0 
\label{EQ:EL1} \\
4 \xi^2 \nabla^4 a_2 - 8 \xi^2 [ ({\bf G}_2 \cdot \nabla)^2 + 
({\bf G}_3 \cdot \nabla)^2 ] a_2 & + & v_2 ( a_1, a_2 ) 
\nonumber \\ 
& = & 0,
\label{EQ:EL2}
\eea
where
%
%
\bea
v_1 (a_1, a_2) & = & 2 \tau a_1 - 2 \gamma a_2^2 + a_1^3 + 4 a_1 a_2^2 
\nonumber \\
v_2 (a_1, a_2) & = & 4 \tau a_2 - 4 \gamma a_1 a_2 + 6 a_2^3 + 4 a_1^2 a_2. 
\label{EQ:V1V2}
\eea
The free-energy (\ref{EQ:LBSINGLE}) and the amplitude equations (\ref{EQ:EL1})
and (\ref{EQ:EL2}) form the basis for our analysis of the droplet nucleation
problem.
%
%
\subsection{Equilibrium phase diagram}

Uniform, periodic lamellar and cylindrical states are position-independent
solutions of Eqs.\ (\ref{EQ:EL1}) and (\ref{EQ:EL2}). When $a_1 = a_2 = 0$, 
the system is disordered. The uniform lamellar solution, 
which exists for $\tau < 0$, is
%
%
\bea
a_1 & = & a_l \equiv  \sqrt{-2 \tau} 
\label{EQ:LAM1} \\
a_2 & = & 0.
\label{EQ:LAM2}
\eea
From Eq.\ (\ref{EQ:LBSINGLE}) we see the uniform lamellar solution has 
free-energy
%
%
\be
f_l = - \tau^2.
\label{EQ:LAMFE}
\ee
The uniform cylindrical solution has $a_1 = a_2 = a_c$ with
%
%
\be
a_c = \frac{\gamma \pm \sqrt{\gamma^2 - 10 \tau}}{5},
\label{EQ:CYLAMP}
\ee
and has a free-energy
%
%
\be
f_c = 3 \tau a_c^2 - 2 \gamma a_c^3 + \frac{15}{4} a_c^4.
\label{EQ:CYLFE}
\ee
For  $\gamma  >  0$  the  solution  with the  positive  root  in  Eq.\
(\ref{EQ:CYLAMP})  has  the  lower  free  energy  and  corresponds  to
cylinders  of A  in a  B matrix.   The case  for $\gamma  < 0$  can be
constructed  from  the $\gamma  >  0$  case  by recognizing  that  the
free-energy,  Eq.\   (\ref{EQ:CYLFE}),  is  invariant   under  $\gamma
\rightarrow -\gamma$ and $a_c \rightarrow -a_c$. It is straightforward
to show that, along a  phase boundary or spinodal, the relations $\tau
=  x \gamma^2$ and  $a_c =  \gamma \tilde{a}_c$  hold for  $\gamma$ of
either sign, with
%
%
\be
\tilde{a}_c = \frac{1 + \sqrt{1 - 10 x}}{5}
\label{EQ:ACT}
\ee
and $x$ a constant. The lamellar/cylinder phase boundary occurs for
%
%
\be
x = - \frac{7 + 3 \sqrt{6}}{5} = -2.8697,
\label{EQ:X}
\ee
as noted in \cite{TURNER94},  and the cylinder/disorder phase boundary
occurs for $x = 4/45$.  The stability limits, or spinodals, for these
uniform phases  can be found by expanding  Eq.\ (\ref{EQ:LBSINGLE}) to
second-order in  deviations from the uniform solution  and looking for
parameters for  which the matrix  of partial second-derivatives  has a
negative eigenvalue.  The disorder-to-order  spinodal is just  $\tau =
0$. On increasing temperature (increasing $\tau$, decreasing $\chi N$) 
the stability  limit for the lamellar
phase is reached when
%
%
\be
\tau = - 2 \gamma^2.
\ee
At this point the lamellar structure will spontaneously transform into the
cylinder phase. 
The spinodal for the reverse,  cylinder-to-lamellar, transition occurs when
$\tau = -8 \gamma^2$.
%
%
\begin{figure}
\resizebox{3.25in}{2.43in}{\includegraphics{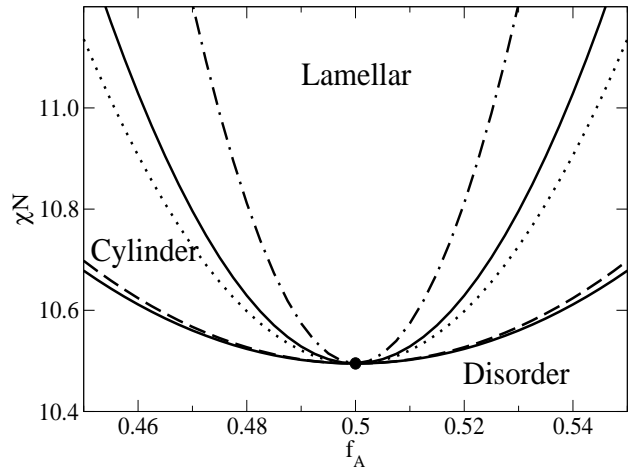} }
\caption{
Equilibrium phase diagram  for the Landau-Brazovskii model in
the  single-mode approximation,  Eq.\  (\ref{EQ:LBSINGLE}). The  solid
lines are  phase boundaries between  the indicated phases.  The dashed
line  is  the disorder-to-order  spinodal.  The dotted  line  is  the
lamellar-to-cylinder   spinodal.  The   dashed-dotted   line  is   the
cylinder-to-lamellar spinodal. The body-centred-cubic spherical phase,
which is stable between the cylinder and disordered phases, and
the gyroid  phase, which  self-consistent field-theory predicts  is
stable for $\chi N  > 11.14$  along the  lamellar/cylinder phase
boundary,   are  not   shown.  The
second-order, mean-field critical point is indicated by the dot. }
\label{FIG:EQPHASE}
\end{figure}

We construct  the equilibrium phase  diagram for the  free-energy Eq.\
(\ref{EQ:LBSINGLE}) in  Fig.\ \ref{FIG:EQPHASE}  in terms of  $\chi N$
and      $f_A$,     using     relations      (\ref{EQ:TAUCHI})     and
(\ref{EQ:GAMVERT}).  In mean-field theory there is a second-order 
critical point at $f_A = 0.5$ and $\chi N = 10.495$ ($\gamma = \tau = 0$),
as indicated.
In self-consistent field-theory  calculations the
body-centred-cubic spherical phase is stable between the
cylindrical and disordered phases and the gyroid phase is stable along
the lamellar/cylinder phase  boundary for $\chi N >  11.14$ and $f_A <
0.452$  $(f_A > 0.548)$  \cite{MATSEN94}. We  have not  included these
phases  in  our  diagram  since   our  primary  concern  is  with  the
lamellar/cylinder  transition and since we  work near  the lamellar/cylinder
phase  boundary  for  $f_A  \geq  0.45$.  As  mentioned  previously,  the
free-energies  of   the  equilibrium  phases   calculated  from  Eq.\
(\ref{EQ:LBSINGLE})  are  within a  few  percent of  those obtained
from self-consistent field-theory.
%
%
\subsection{Planar interface between the lamellar and cylindrical phases}

For a planar interface  separating the lamellar and cylindrical phases
the amplitudes  $a_1$ and $a_2$ become functions  of the perpendicular
distance  to  the interface,  $s  = {\bf  \hat{n}}  \cdot  {\bf r  }$,
only. The unit normal to the  interface is ${\bf \hat{n}} = (\cos \phi
\sin \theta, \sin \phi \sin  \theta, \cos \theta)$, expressed in terms
of  the angles  defined in  Fig. \ref{FIG:GEOMETRY}.   From Eq.\
(\ref{EQ:LBSINGLE}),  the  free-energy   per  unit  interfacial  area,
$\tilde{f}$, is then
%
%
\bea
\tilde{f} & = & \int ds \Big\{ \xi^2 
(a''_1)^2 + 2 \xi^2 (a''_2)^2 
+ 4 \xi^2 [ ({\bf G}_1 \cdot {\bf \hat{n}})^2 (a'_1)^2 
\nonumber \\ & & 
+ \{ ({\bf G}_2 \cdot {\bf \hat{n}})^2 + 
({\bf G}_3 \cdot {\bf \hat{n}})^2 \} (a'_2)^2 ] 
+ \tau (a_1^2 + 2 a_2^2)
\nonumber \\ & &
- 2 \gamma a_1 a_2^2 + \frac{1}{4} ( a_1^4 + 6 a_2^4 + 8 a_1^2 a_2^2 ) \Big\},
\label{EQ:1DFE}
\eea
where primes denote differentiation with respect to $s$. We note that 
%
%
\bea
({\bf \hat{n}} \cdot {\bf G}_1)^2 & = &  \sin^2 \phi \mbox{ } \sin^2 \theta \\
({\bf \hat{n}} \cdot {\bf G}_2)^2 + ({\bf \hat{n}} \cdot {\bf G}_3)^2 
& = & \frac{1}{2} (1 + 2 \cos^2 \phi ) \sin^2 \theta.
\eea
Thus terms involving the squares of single derivatives of the field 
(square-gradient terms) do
not  contribute to  $\tilde{f}$ when  the interface  normal is  in the
$\hat{\bf z}$ direction ($\theta = 0$).
%
%
\section{Energetics of the lamellar/cylinder interface}
\label{SEC:ANSATZ}

The aim of this section  is to compute the interfacial free-energy for
a  planar interface  of  arbitrary orientation  $\hat{\bf n}$  between
coexisting  lamellar  and  cylindrical  phases. Once  the  interfacial
free-energy  is known,  the Wulff  construction \cite{WULFF01}  can be
used to find the minimal-energy droplet  shape, as we show in the next
section. It  is sufficient to approximate  the interfacial free-energy
near  the lamellar/cylinder phase  boundary by  its value  at boundary
since the  interfacial free-energy function is  continuous through the
transition.
%
%
\subsection{An analytic calculation of the interfacial free-energy}

Rather than numerically solving the Euler-Lagrange equations resulting
from  Eq.\ (\ref{EQ:1DFE})  to get  the interfacial  profile  and then
computing the interfacial free-energy, we instead employ a variational
ansatz  for  the  interfacial  profile which  allows  the  interfacial
free-energy to be calculated analytically. Our variational solution to
the problem will, of  course, produce an interfacial free-energy which
is  greater  than that  obtained  through  an  exact solution  of  the
Euler-Lagrange equations.  However, our analytical  solution will give
us  much insight  into the  behaviour of  the  interfacial free-energy
along the lamellar/cylinder phase boundary.

We employ  the following variational ansatz for  the amplitude profile
of a planar interface between coexisting cylinder and lamellar phases,
%
%
\bea
a_1(s) & = & \left( \frac{a_l + a_c}{2} \right) 
+ \left( \frac{a_l - a_c}{2} \right) h \left( \frac{s}{w} \right) 
\label{EQ:ANSATZ1} \\
a_2(s) & = & \frac{a_c}{2} \left[ 1 -  h \left( \frac{s}{w} \right) \right]. 
\label{EQ:ANSATZ2}
\eea
The function $h$  has the properties that $h(s)  \rightarrow \pm 1$ as
$s \rightarrow  \pm \infty$ and $h(0)  = 0$, thus it  describes a pure
cylinder phase for $s \rightarrow  - \infty$ separated by an interface
at  $s=0$ from  a  pure  lamellar phase,  obtained  as $s  \rightarrow
\infty$. The  form of  $h$ will be  specified later.   The variational
parameter, $w$,  characterizes the interfacial width,  and will depend
on    the    interface    orientation.     Substitution    of    Eqs.\
(\ref{EQ:ANSATZ1}) and  (\ref{EQ:ANSATZ2}) into the  free-energy, Eq.\
(\ref{EQ:1DFE}), and  rescaling $s$ to extract factors  of $w$ results
in the excess free-energy due to the interface,
%
%
\be
\tilde{f} - \tilde{f}_{h} = \frac{\xi^2 \gamma^2 g_1}{w^3} 
+ \frac{\xi^2 \gamma^2 g_2}{w} + \gamma^4 g_3 \mbox{ } w,
\label{EQ:FEW}
\ee
where 
%
%
\be  \tilde{f}_{h} =  \int  ds \mbox{  }  f_l \ee  is the  free-energy
contribution  from  the  two  uniform  phases  ($f_l  =  f_c$  at  the
lamellar/cylinder   phase  boundary).    The   coefficients  in   Eq.\
(\ref{EQ:FEW}) are
%
%
\bea
g_1  & = & \frac{1}{4} I_1 (\tilde{a}_l^2 - 2 
\tilde{a}_l \tilde{a}_c + 3 \tilde{a}_c^2)  \\
g_2  & = & \frac{1}{2} I_2  
[ 3 \tilde{a}_c^2 + 2 \tilde{a}_l ( \tilde{a}_l - 2 \tilde{a}_c) \sin^2 \phi 
]\sin^2 \theta  
\label{EQ:G2} \\
g_3 & = & \int_{-\infty}^{\infty} 
ds \Big\{ x (\tilde{a}_1^2 + 2 \tilde{a}_2^2) - 2  \tilde{a}_1 \tilde{a}_2^2 
\nonumber \\ & & 
+ \frac{1}{4} ( \tilde{a}_1^4 + 6 \tilde{a}_2^4 + 8 \tilde{a}_1^2 
\tilde{a}_2^2 ) 
+ x^2 \Big\},
\label{EQ:G3CL}
\eea
where
\bea
I_1 & = & \int_{-\infty}^{\infty} ds \mbox{ } (h'')^2 \\
I_2 & = & \int_{-\infty}^{\infty} ds \mbox{ } (h')^2.
\eea
The quantity  $\tilde{a}_l = \sqrt{-2  x}$, and $\tilde{a}_c$  and $x$
are  given   by  Eqs.\  (\ref{EQ:ACT})  and   (\ref{EQ:X}).   In  Eq.\
(\ref{EQ:G3CL})  the  functions  $\tilde{a}_1$ and  $\tilde{a}_2$  are
given  by Eqs.\ (\ref{EQ:ANSATZ1})  and (\ref{EQ:ANSATZ2})  with $a_l$
and  $a_c$ replaced by  $\tilde{a}_l$ and  $\tilde{a}_c$ respectively,
and  with  $w=1$.   The  coefficients  $g_1$,  $g_2$,  and  $g_3$  are
positive.   The excess free-energy,  Eq. (\ref{EQ:FEW}),  is minimized
when
\be
w^2 = (w^*)^2 \equiv \frac{\xi^2 g_2 + \sqrt{ \xi^4 g_2^2 +
 12 \xi^2 \gamma^2 g_1 g_3 }}{2 \gamma^2 g_3}.
\label{EQ:WM}
\ee
Substitution of this value for $w$ in  Eq.\ (\ref{EQ:FEW}) provides the 
interfacial free-energy, $\sigma(\theta,\phi) \equiv \tilde{f}^* - 
\tilde{f}_{h}$, for a planar cylinder/lamellar interface with normal  
$\hat{\bf n}$,
%
%
\be
\sigma(\theta,\phi)
= \frac{\xi^2 \gamma^2 g_1}{(w^*)^3} + \frac{\xi^2 \gamma^2 g_2}{w^*} + 
\gamma^4 g_3 w^*.
\label{EQ:STW}
\ee
The angular  dependence in Eq.\  (\ref{EQ:STW}) is contained  in $g_2$
and  $w^*$.   Equation (\ref{EQ:G2})  indicates  that the  interfacial
free-energy  depends on the  polar angle  $\theta$ through  factors of
$\sin^2   \theta$.   Equation  (\ref{EQ:G2})   also  shows   that  the
interfacial free-energy is invariant  under $\phi \rightarrow \phi + n
\pi$, where  $n$ is an  integer (two-fold azimuthal symmetry).

To  examine the  behaviour  of $\sigma(\theta,\phi)$  along the  phase
boundary  near the mean-field 
critical  point, where  $\gamma \rightarrow  0$, we
have  to distinguish  between  two cases,  depending  on the  relative
values of $\theta$  and $\gamma$.  For $|\gamma| \ll  g_2$, as obtains
for finite $\theta$ and $\gamma \rightarrow 0$, we have
%
%
\be
w^* = \frac{\xi}{|\gamma|} \sqrt{ \frac{g_2}{g_3} } + O(|\gamma|)
\label{EQ:WCRIT}
\ee
and
\be
\sigma(\theta,\phi) = 2 \xi |\gamma|^3 \sqrt{g_2 g_3} + O(|\gamma|^5).
\label{EQ:STCRIT}
\ee
For $\theta \rightarrow 0$ such that $ g_2 \ll |\gamma|$ we have
%
%
\bea
w^* & = & \left( \frac{\xi}{|\gamma|} \right)^{1/2} \left( 
\frac{3 g_1}{g_3} \right)^{1/4} \left[ 1 + 
\frac{\xi g_2}{2 |\gamma| (12 g_1 g_3)^{1/2}} \right.
\nonumber \\ & & \hspace{1in} \left.
+ O \left( \frac{g_2^2}{\gamma^2} \right) \right]
\label{EQ:WCRIT0}
\eea
and
\bea
\sigma (\theta,\phi) & = & \frac{4}{3} \xi^{1/2} |\gamma|^{7/2} ( 3 g_1 g_3^3
)^{1/4} \left[ 1 + \frac{3 \xi g_2}{2 | \gamma | (12 g_1 g_3)^{1/2}} 
\right. \nonumber \\ & & \hspace{1in} \left.
+ O \left( \frac{g_2^2}{\gamma^2} \right) \right].
\label{EQ:STCRIT0}
\eea
The width  of the  interface diverges at  the mean-field
critical point,  and the
interfacial  free-energy vanishes  there.  It  is interesting  to note
that  the  interfacial  free-energy  remains  anisotropic  as  $\gamma
\rightarrow  0$.  Since  $\gamma  \sim \sqrt{\tau}$  along  the  phase
boundary, the scaling with $\gamma$ seen in Eqs.\ (\ref{EQ:WCRIT}) and
(\ref{EQ:STCRIT}) corresponds  to the expected  mean-field results, $w
\sim \tau^{-1/2}$  and $\sigma  \sim \tau^{3/2}$ \cite{STMFT}  . When
$g_2 \ll |\gamma|$ Eqs.\ (\ref{EQ:WCRIT0}) and (\ref{EQ:STCRIT0}) show
that, as  $\gamma \rightarrow 0$, the interfacial  width diverges more
slowly  than the  mean-field  result and  the interfacial  free-energy
vanishes  more   rapidly  than  the  mean-field   result.  As  $\gamma
\rightarrow 0$, the condition  $g_2 \ll |\gamma|$ restricts the region
over  which  this  behaviour   holds  to  very  small  $\theta$.   The
non-mean-field     scaling    in    Eqs.\     (\ref{EQ:WCRIT0})    and
(\ref{EQ:STCRIT0})  arises  because,  for   $\theta  =  0$,  only  the
square-second-derivative    terms   in   Eq.\    (\ref{EQ:1DFE})   are
non-vanishing.   Finally, far  from the  critical point,  as $|\gamma|
\rightarrow  \infty$, the  condition $g_2  \ll \gamma$  holds  for all
$\theta$ and the  expressions (\ref{EQ:WCRIT0}) and (\ref{EQ:STCRIT0})
apply.  At  leading order $w^* \rightarrow 0$  and $\sigma \rightarrow
\infty$ as $|\gamma| \rightarrow \infty$.  The interfacial free-energy
becomes isotropic in this limit.

None  of these  results  depend on  the  form chosen  for  $h$ in  the
variational   approximation.   The   only   approximation   in   Eqs.\
(\ref{EQ:ANSATZ1})  and  (\ref{EQ:ANSATZ2})  is that  the  interfacial
profiles  for two different  orientations have  the same  basic shape,
$h$, and  can be scaled  onto one another by  an orientation-dependent
scale-factor $w$.
%
%
\subsection{Results for the lamellar/cylinder interfacial free-energy}

To calculate  the interfacial free-energy  we will now choose  $h(s) =
\tanh  s$, which  is a  reasonable approximation  for  an interfacial
profile, and satisfies the  boundary conditions mentioned after Eqs.\
(\ref{EQ:ANSATZ1})  and (\ref{EQ:ANSATZ2}). With  this choice  $I_1 =
16/15$, $I_2 =  4/3$ and $g_3 \approx$ 0.716  286. The unscaled ({\em
i.e.}\  actual) interfacial  free-energy, $\sigma_0$,  is  related to
$\sigma$ in Eq.\ (\ref{EQ:STW}) by
%
%
\be
\sigma_0 = \frac{\lambda_0}{\sqrt{6 x^*}} \left( \frac{\rho_0 \mbox{ } b 
\mbox{ } k_B T}{N^{1/2}} 
\right) \mbox{ } \sigma.
\label{EQ:ST0}
\ee
%
%
\begin{figure}[h]
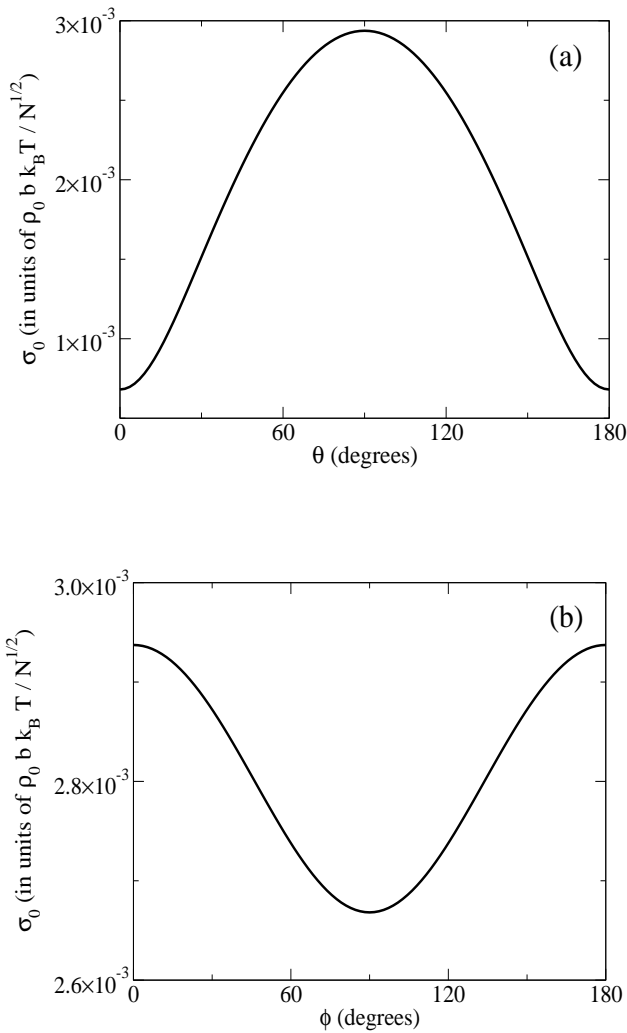

\resizebox{3.25in}{2.43in}{\includegraphics{figure3a.eps} }
\\
\vspace{0.5 in}
\resizebox{3.25in}{2.43in}{\includegraphics{figure3b.eps} }
\vspace{.33 in}
\caption{ Interfacial  free-energy, $\sigma_0$  in units of  $\rho_0 \mbox{ } 
b \mbox{ } 
k_B  T   /  N^{1/2}$,  of  the  lamellar/cylinder   interface  at  the
lamellar/cylinder phase boundary  for $f_A = 0.45$ ($\chi  N = 11.3263$)
as per Eqs.\  (\ref{EQ:STW}) and (\ref{EQ:ST0}). (a) As  a function of
$\theta$, for  $\phi = 0$. (b) As  a function of $\phi$  for $\theta =
\pi /2$  (in the $x$--$y$ plane).  The angles $\theta$  and $\phi$ are
defined in Fig.\ \ref{FIG:GEOMETRY}.  }
\label{FIG:ST}
\end{figure}

In  Fig.\ \ref{FIG:ST}  we present results  for $\sigma_0$  at the
phase  boundary,  in  units  of  $\rho_0  \mbox{  }  b  \mbox{  }  k_B
T/N^{1/2}$,  for  $f_A  =  0.45$.  Figure  \ref{FIG:ST}  qualitatively
represents the  features of the interfacial free-energy  for all $f_A$
near  the mean-field  critical point.   The overall  magnitude  of the
interfacial free-energy increases with stronger segregation.  In Fig.\
\ref{FIG:ST}a we show $\sigma_0$ as a function of $\theta$ for $\phi =
0$.  
When the cylinder axis is perpendicular to the interface ($\theta
= 0$) the  interfacial free-energy is about 4  times smaller than when
the cylinders  are parallel to  the interface ($\theta =  \pi/2$).  In
Fig.\  \ref{FIG:ST}b  the interfacial  free-energy  as  a function  of
$\phi$  in  the $x$--$y$  plane  ($\theta  =  \pi/2$) is  shown.   The
interfacial free-energy is highest when the lamellae lie perpendicular
to  the  interface ($\phi  =  0$) and  lowest  when  the lamellae  are
parallel  to the  interface ($\phi  = \pi/2$).   The variation  of the
interfacial free-energy  with $\phi$ is  less than its  variation with
$\theta$,  since $2  \tilde{a}_l(\tilde{a}_l  - 2  \tilde{a}_c) \ll  3
\tilde{a}_c^2$  in   Eq.\  (\ref{EQ:G2}).  Although   the  interfacial
free-energies for  $\phi = \pi/6$ and  $\phi = 5 \pi/6$  are equal, as
they  should  be, it  is  somewhat  surprising  that the  presence  of
cylinder lattice planes at these  angles does not produce any features in the
interfacial free-energy. It may be that at higher segregation, where 
one has to go beyond the single-mode approximation, the existence of 
cylinder lattice planes will become manifest in the droplet shape.
Similar work by Netz {\em et al.}\ examined the lamellar/cylinder
interfacial free-energy numerically when $\theta = \phi =  \pi/2$
\cite{NETZ97}.  They  found the mean-field scaling, $\sigma_0
\sim \tau^{3/2}$, near  the critical point in their  model. Self-consistent   
field-theory has been used recently to study the energetics
of kink-  and twist-grain boundaries in diblock copolymer melts --- 
situations also involving complicated interfacial structures 
\cite{MATSEN97,DUQUE00}.
%
%
\section{Wulff Construction of the Critical Droplet}
\label{SEC:WULFF}

In  general,  the  nucleating  droplet  of cylindrical  phase  is  not
spherical and it is necessary  to calculate the droplet shape from the
anisotropic  interfacial  free-energy,  using the  Wulff  construction
\cite{WULFF01}. Once  the droplet shape  is known the  free-energy and
size of the critical droplet can be calculated.

\subsection{Wulff construction for the droplet shape}

The  droplet  shape  is   found  by  minimizing  the  droplet  surface
free-energy,   subject   to  the   constraint   of  constant   droplet
volume. Thus we minimize the function
%
%
\be
F_{Wulff} =  S_{drop} - 2 \mu V_{drop},
\label{EQ:WULFF}
\ee 
where $\mu$ is a  Lagrange multiplier, $V_{drop}$ is the fixed
droplet volume and
%
%
\be
S_{drop} \equiv \int dS \mbox{ } \sigma (\hat{{\bf n}}) 
\label{EQ:SDROP}
\ee
is the  integral of the interfacial  free-energy, Eq.\ (\ref{EQ:STW}),
over the  droplet surface.  Implicit  in Eq.\ (\ref{EQ:WULFF})  and in
the classical  nucleation theory we employ is  the assumption that
the  droplet interface  width is  negligible compared  to  the droplet
size, so that a separation may be made between the bulk and surface of
the droplet.  We will  show this  to be the  case {\em  a posteriori},
close to the lamellar/cylinder phase boundary.

The minimization  of Eq.\  (\ref{EQ:WULFF}) is performed  by choosing,
for example,  the $x$--$y$ plane  and characterizing the shape  of the
droplet by the height  of the interface, $h_0({\bf x}_{\perp})$, above
this plane at  ${\bf x}_{\perp} = (x,y)$.  Minimization  then leads to
the following formula for the height \cite{CHAIKIN95},
%
%
\be
\tilde{h}(\tilde{\bf x}_{\perp}) = [ \tilde{g} ({\bf m}) + 
{\bf m} \cdot {\bf  \tilde{x}}_{\perp} 
]_{\mbox{min {\bf m}}},
\label{EQ:HEIGHT}
\ee
where
%
%
\bea
\tilde{h} & = & \frac{h_0}{L} \\
{\bf \tilde{x}}_{\perp} & = & \frac{{\bf x}_{\perp}}{L} \\
{\bf m} & = & \nabla_{\perp} h_0 \\
 \tilde{g} ({\bf m}) & = & (1 + {\bf m}^2 )^{1/2} \frac{\sigma [
\hat{\bf n}({\bf m})]}
{\sigma_{max}}.
\label{EQ:FDEF}
\eea 
We have written $\mu  = \sigma_{max}/ L$ with $\sigma_{max} =
[\sigma  ({\bf\hat{n}}({\bf  m}))]_{\mbox{max  {\bf  m}}}$ and  $L$  a
length-scale determined  by the value chosen  for $V_{drop}$. Equation
(\ref{EQ:HEIGHT})  is  the essence  of  the  Wulff construction.   The
minimization in Eq.\ (\ref{EQ:HEIGHT}) over the variable ${\bf m}$ for
a given ${\bf \tilde{x}}_{\perp}$ is performed numerically.
%
%
\begin{figure}
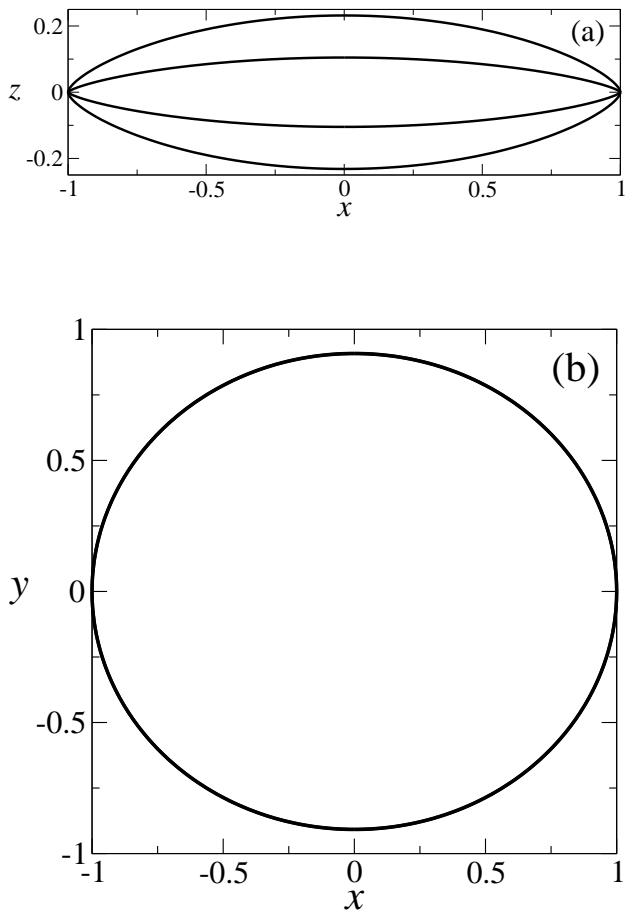

\resizebox{3.25in}{!}{\includegraphics{figure4a.eps} }
\vspace{0.5 in} \\
\resizebox{3.25in}{!}{\includegraphics{figure4b.eps} }
\vspace{.33 in}
\caption{ 
Cross-section of the  droplet shape calculated {\em via} the
Wulff construction. We have chosen the unit of length to be the largest 
dimension of 
the droplet, in this case the half-length along  the  
$x$-axis. (a) Cross-section
in the  $x$--$z$ plane for $f_A  = 0.45$ (outermost curve)  and $f_A =
0.49$  (innermost curve) along  the lamellar/cylinder  phase boundary.
The  cylinders inside  the  droplet are  oriented  along the  $z$-axis
(vertically) and the lamellae outside  the droplet lie in the plane of
the page.  (b) Cross-section  in the $x$--$y$  plane for $f_A  = 0.45$
(outermost  curve)  and  $f_A  =  0.49$ (innermost  curve)  along  the
lamellar/cylinder   phase   boundary.    These   curves   are   essentially
indistinguishable. The cylinders inside the droplet are in a hexagonal
lattice,  with their  axes  pointing  out of  the  page. The  lamellae
outside  the droplet  are  oriented horizontally,  with their  normals
along the $y$-axis.}
\label{FIG:2DSHAPE}
\end{figure}

The droplet shape along  the lamellar/cylinder phase boundary for $f_A
= 0.45$ and $0.49$ is  shown in Fig.\ \ref{FIG:2DSHAPE}. 
In this figure we have chosen the unit of length to be the largest 
dimension of 
the droplet, in this case the half-length, $l_x$,  along  the  
$x$-axis. The other dimensions are 
determined, in these units, through the Wulff construction.
  Figure \ref{FIG:2DSHAPE}a  shows that  the
droplet is lens-shaped, being flattened  along the axis of the cylinders
($z$-axis). This  follows from  the lower interfacial  free-energy for
interfaces  perpendicular  to the  cylinders,  compared to  interfaces
parallel to the cylinders. The  anisotropy of the droplet increases as
the  mean-field critical  point   is  approached. 
The  droplet  shape
anisotropy can be characterized by  the ratio of $l_x$  to  the
half-length  along  the  $z$-axis,
$l_z$.  From Eqs.\ (\ref{EQ:HEIGHT})  and (\ref{EQ:FDEF})  this aspect
ratio is
%
%
\be
\frac{l_x}{l_z} = \frac{\sigma(\pi/2,0)}{\sigma(0,0)},
\label{EQ:ASPECTZX}
\ee
which, from Eqs. (\ref{EQ:STCRIT}) and  (\ref{EQ:STCRIT0}), scales as 
%
%
\be 
\frac{l_x}{l_z} \sim \xi^{1/2} \mbox{ } |\gamma|^{-1/2} 
\ee 
as the critical  point   is  approached,   and  approaches  1   as  
$|\gamma| \rightarrow \infty$. The increase in the aspect ratio of the
droplet as the mean-field critical point is approached is is a 
consequence of the gradient structure of the theory, which leads to different 
scaling of the interfacial free-energy with $|\gamma|$ in 
Eqs.\ (\ref{EQ:STCRIT}) and (\ref{EQ:STCRIT0}).

Figure  \ref{FIG:2DSHAPE}b shows  that  the droplet  is also  slightly
flattened along the $y$-axis,  compared to the $x$-axis. This reflects
the  trend   seen  in  Fig.\  \ref{FIG:ST}b,   where  the  interfacial
free-energy is  lower when the  lamellae lie along the  interface than
when  they are  perpendicular to  it. As  expected, the  droplet shape
anisotropy is weaker in the $x$--$y$  plane than it is in the $x$--$z$
plane.  The change  in the  droplet shape  anisotropy in  the $x$--$y$
plane with $f_A$ is also very  small. If $l_y$ is the half-length of the
droplet along the  $y$-axis, Eqs.\ (\ref{EQ:HEIGHT}), (\ref{EQ:FDEF}),
and (\ref{EQ:STCRIT}) lead to
%
%
\be
\frac{l_y}{l_x} = \sqrt{\frac{g_2(\pi/2,\pi/2)}{g_2(\pi/2,0)}} = 0.9074,
\ee
at   leading  order   in   $|\gamma|$  as   the   critical  point   is
approached.  Thus near  the  mean-field 
critical point  the  droplet will  become
extremely flattened  along the $z$-axis, with  a slightly non-circular
cross section in the $x$--$y$ plane. As $|\gamma| \rightarrow \infty$,
Eq.\  (\ref{EQ:STCRIT0})  indicates   that  the  droplet shape  will  become
isotropic,  and $l_y/l_x  \rightarrow 1$.  As anticipated,  
the presence of cylindrical lattice planes at $\phi = \pi/6$ and 
$\phi = 5 \pi /6$ is not manifest in the droplet shape in the $x$--$y$
plane.  We  will  compare these  results for  the
droplet shape with relevant  experimental and numerical results in the
Discussion section.
%
%
\subsection{The critical droplet}

The  droplet  volume  serves  as  a  reaction  coordinate  during  the
transformation  from  the  metastable  lamellar phase  to  the  stable
cylinder phase. Once  this volume is selected, the  droplet shape that
minimizes  the  surface  free-energy  is  determined  from  the  Wulff
construction.  We  will employ  the  classical homogeneous  nucleation
theory \cite{NUCLEATION}, in which the critical droplet corresponds to
a total free-energy maximum  along the volume reaction coordinate, and
the total free-energy is a sum of surface and bulk terms. The computed
droplet  shape is  independent of  the  value chosen  for the  droplet
volume,  thus  the shape  of  the  critical  droplet  will  be  just  that
calculated in the last section.

When we choose  the unit of length to be the  largest dimension of the
droplet, as  in the last section,  we can compute  the droplet volume,
$V_{drop}$,  and surface  free-energy, $S_{drop}$,  in these  units as
functions of $f_A$.  When $f_A = 0.45$, for example, $S_{drop}=$ 1.949
57  $\times  10^{-4} $  and  $V_{drop}=$  0.784  000 (the  interfacial
free-energy in  $S_{drop}$, Eq.\ (\ref{EQ:SDROP}),  is dimensionless).
If  we wish  to  express $l_x$  in  units of  $q_0^{-1}$, the  surface
free-energy  $S(l_x)$  and  droplet   volume  $V(l_x)$,  in  units  of
$q_0^{-2}$ and $q_0^{-3}$ respectively, will scale as
%
%
\bea   
S(l_x)  &   =   &   S_{drop} \mbox{ }   l_x^2  \\   
V(l_x)  &   =  &   V_{drop} \mbox{ } l_x^3 
\eea   
for arbitrary $l_x$. 

We  now examine  the system  slightly  away from  the phase  boundary,
making the standard assumption of classical nucleation theory that the
interfacial   free-energy  and  interfacial   width   do   not  change
significantly  near the  phase boundary.  When the  lamellar  phase is
metastable and the cylinder phase is stable we have
%
%
\be
\Delta f \equiv f_c - f_l < 0.
\ee
Separating the  free-energy $F_0$ (in real free-energy units)
into a bulk and a surface term, and subtracting the free-energy, $F_l$, of the
metastable  uniform lamellar  phase we  have
%
%
\be
\frac{(F_0 - F_l) N}{\rho_0 k_B T} = \frac{\lambda_0}{q_0^3} 
\left[ V_{drop} \mbox{ } l_x^3 \times\Delta f + 
S_{drop} \mbox{ } l_x^2 \right].
\ee
The  critical droplet occurs  for the half-length, $l_{cx}$, that
maximizes this expression, namely
%
%
\be
l_{cx} =  - \frac{2 S_{drop}}{3 V_{drop} \Delta f},
\label{EQ:RCRIT}
\ee
for which the nucleation barrier, $\Delta F_c \equiv
F_{crit} - F_l$, is given by
\be
\frac{\Delta F_c}{ \rho_0 b^3 N^{1/2} k_B T} = 
\frac{\lambda_0}{(6 x^*)^{3/2}} 
\times \frac{4 (S_{drop})^3}{27 (V_{drop})^2 (\Delta f)^2}.
\label{EQ:NUCL}
\ee
The critical half-length in Eq.\ (\ref{EQ:RCRIT}) 
is expressed in units of $q_0^{-1}$. Equation (\ref{EQ:Q0}) has been used in 
Eq.\ (\ref{EQ:NUCL}).
%
%
\begin{figure}
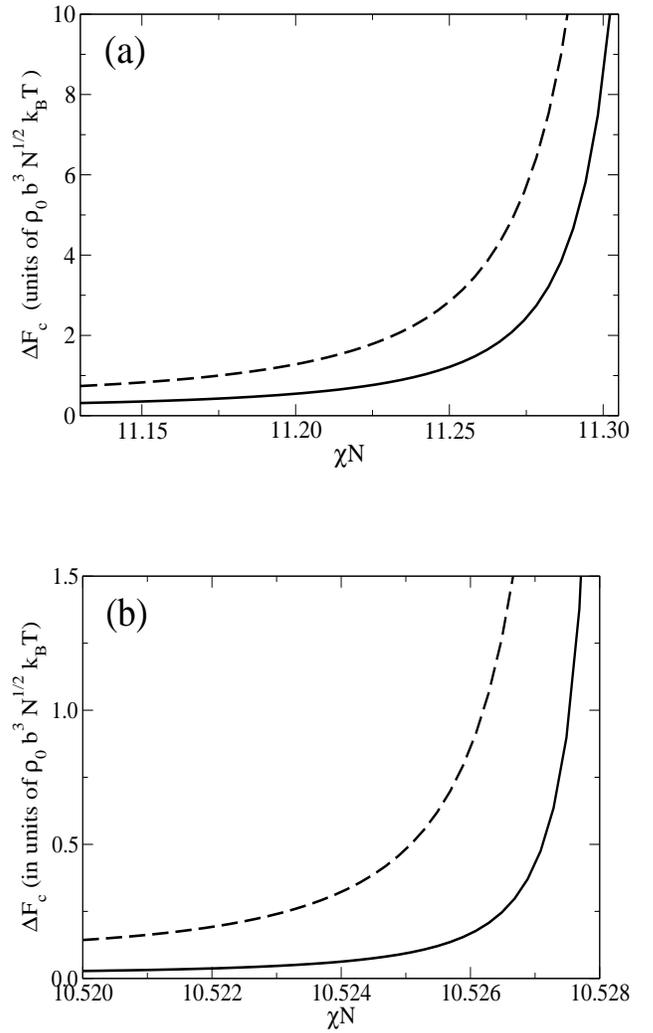

\resizebox{3.25in}{2.43in}{\includegraphics{figure5a.eps} }
\\ \vspace{.5in}
\resizebox{3.25in}{2.43in}{\includegraphics{figure5b.eps} } 
 \\
\caption{ Nucleation barrier (solid curves)  as a function of $\chi N$
near  the lamellar/cylinder  phase boundary.   (a) For  $f_A  = 0.45$,
where the phase boundary occurs at $\chi N = 11.3263$ and the spinodal
occurs at  $\chi N = 11.1358$. (b)  For $f_A = 0.49$,  where the phase
boundary occurs at $\chi N = 10.5285$ and the spinodal occurs at $\chi
N  = 10.5207$. For  comparison, the  dashed curves  are the  result of
using  the anisotropic  interfacial free-energy,  Eq.\ (\ref{EQ:STW}),
but assuming a spherical droplet shape.}
\label{FIG:BARRIER}
\end{figure}

In Fig.\  \ref{FIG:BARRIER} we plot  $\Delta F_c$ in units  of $\rho_0
b^3 N^{1/2} k_B T$ near  the lamellar/cylinder phase boundary for $f_A
=  0.45$ and  $0.49$.  The  nucleation barrier  diverges at  the phase
boundary,  but remains  finite at  the  spinodal ---  behaviour to  be
expected from classical  nucleation theory. Physically, the nucleation
barrier should  go to  zero at  the spinodal, but  this is  beyond the
classical nucleation approach employed here.  Over the region in $\chi
N$ from  the spinodal to the  phase boundary the general  trend is for
the  nucleation   barrier  to  decrease  as  the   critical  point  is
approached.   The importance  of  Fig.\ \ref{FIG:BARRIER}  is that  it
allows one  to calculate  the magnitude of  the nucleation  barrier in
units of $k_B T$, once $N$,  $\rho_0$ and $b$ are known. Since $\Delta
F_c$ scales  as $N^{1/2}$, we  expect polymers with higher  indices of
polymerization  will  have larger  nucleation  barriers.  Finally,  in
Fig.\ \ref{FIG:BARRIER} we plot  the nucleation barrier obtained using
our  theory for  the  interfacial free-energy,  but  with a  spherical
droplet  shape   (dashed  curve).  In   this  case  $\Delta   F_c$  is
significantly  increased,  indicating  the  importance  of  using  the
proper, anisotropic  droplet shape for an accurate  calculation of the
nucleation barrier.
%
%
\begin{figure}
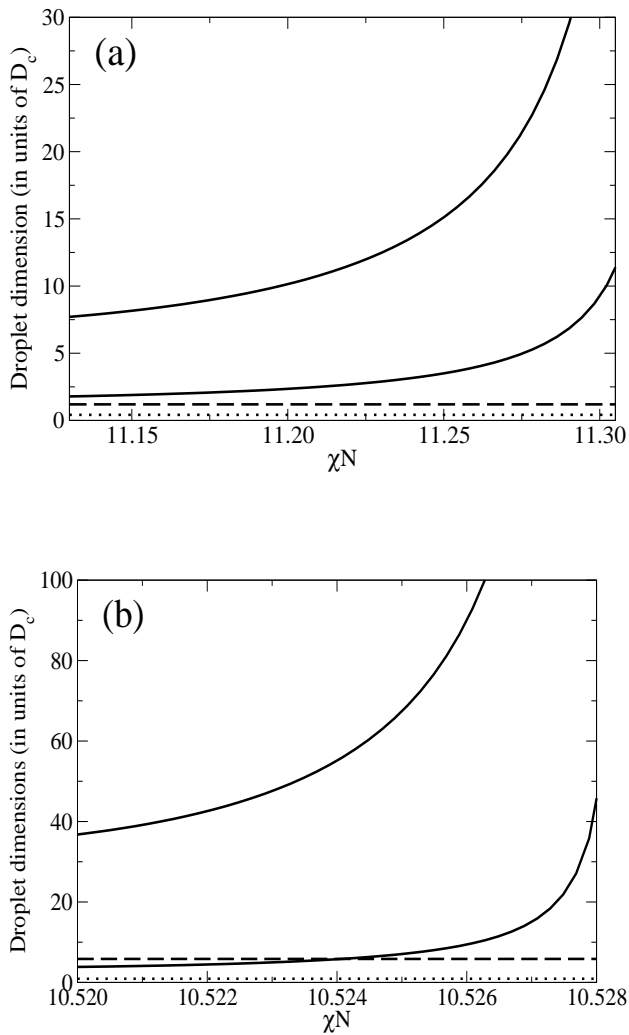

\resizebox{3.25in}{2.43in}{\includegraphics{figure6a.eps} }
\\ \vspace{.5 in}
\resizebox{3.25in}{2.43in}{\includegraphics{figure6b.eps} }
\\
\caption{ Critical droplet dimensions,  in units of the distance $D_c$
between   cylinders,   as   a   function   of  $\chi   N$   near   the
lamellar/cylinder phase boundary.  (a) $f_A  = 0.45$ (b) $f_A = 0.49$.
The upper solid  curve corresponds to $l_{cx}$, the  lower solid curve
to    $l_{cz}$.    The   interface    widths,   calculated    at   the
lamellar/cylinder phase boundary,  are indicated: $w_x$ (dashed line),
$w_z$ (dotted line).  The transitions  and spinodals occur in the same
places as in Fig.\ \ref{FIG:BARRIER}.}
\label{FIG:LENGTHS}
\end{figure}

In Fig.\  \ref{FIG:LENGTHS} we examine the dimensions  of the critical
droplet,    $l_{cx}$    and     $l_{cz}$    obtained    using    Eqs.\
(\ref{EQ:ASPECTZX}) and  (\ref{EQ:RCRIT}), as  a function of  $\chi N$
near the lamellar/cylinder phase boundary  for $f_A = 0.45$ and $f_A =
0.49$. In this figure the lengths are in units of the cylinder spacing
$D_c$.  In addition to $l_{cx}$  and $l_{cz}$ we also show the droplet
interfacial widths, $w_x$ and  $w_z$, obtained from the calculation of
the  interfacial  free-energy at  the  phase  boundary for  interfaces
oriented  with normals  along the  $x$-- and  $z$--axes, respectively.
The  droplet size  diverges as  the phase  boundary is  approached, as
expected from classical  nucleation theory. In Fig.\ \ref{FIG:LENGTHS}
we  examine the  region in  $\chi N$  from the  spinodal to  the phase
boundary.  The general trend is  for the scale of the critical droplet
in  this  region to  increase  as  the  mean-field critical  point  is
approached.   As   anticipated   from   Eqs.\   (\ref{EQ:WCRIT})   and
(\ref{EQ:WCRIT0}) the width of  the droplet interface increases as the
critical  point  is approached.   Netz  {\em  et  al.}\ also  observed
widening  of  the   lamellar/cylinder  interface  as  the  segregation
decreased \cite{NETZ97}.  For these $f_A$ and $\chi N$ we find $l_{cx}
> w_x$ and $l_{cz} > w_z$. For $f_A = 0.49$ there is a region of $\chi
N$ for which $l_{cz} < w_x$.   In the next section we will discuss the
region of validity  of our theory, and the range  of droplet sizes and
nucleation barriers expected in this region.
%
%
\section{Discussion}

To satisfy the assumptions of the Wulff construction and the classical
nucleation theory  we study the region close  to the lamellar/cylinder
phase  boundary,  but  not   so  close  that  nucleation  is  rendered
unobservable due to extremely  high barriers. In small-molecule binary
fluid mixtures, Cahn and Hilliard estimated that a free-energy barrier
below 60 $k_B  T$ should produce an observable  rate of nucleation (at
least  one  nucleation  event  per  day  per  cubic  centimetre,  say)
\cite{CAHN59}. Since the relaxational  dynamics in polymeric fluids is
slower  than  in small-molecule  fluids,  we  expect  that a  somewhat
smaller  free-energy   barrier  is  required  for   nucleation  to  be
observable in our system. However,  in the absence of a kinetic theory
for nucleation  in polymers, we  will take 60  $k_B T$ to be  an upper
limit for  the nucleation  barrier. In the  context of  binary polymer
blends, Binder  \cite{BINDER84} and, more recently,  Wang  \cite{WANG02} have
argued that mean-field theory breaks down when the free-energy barrier
becomes less  than 10  $k_B T$.  Following  these authors,  we take
a nucleation  barrier  of 10  $k_B  T$ to be  the  lower limit  for
applicability of the present theory.

To be consistent with  the slowly-varying amplitude approximation, the
width   of  the   droplet  interface   should  be   larger   than  the
microstructure  period. For  $f_A =  0.45$  we have  $w_x \approx  w_y
\approx  D_c$, while  closer  to the  mean-field  critical point  $w_x
\approx w_y > D_c$.  We typically find $w_z < D_c$ (when $f_A = 0.45$,
$w_z \approx 0.41$ $D_c$, and it  is not until $f_A = 0.492$ that $w_z
=   D_c$).   However,   since   the  variations   in  the   underlying
microstructure are in the  $x$--$y$ plane, relatively rapid variations
of the amplitude  in the $z$-direction should not  affect the validity
of the  slowly-varying amplitude approximation.  For $f_A  = 0.45$ and
$\chi N = 11.25$ with $N  = 1000$ and $\rho_0 = b^{-3}$ the nucleation
barrier is  39 $k_B T$ and  the critical droplet, which  has an aspect
ratio  of  4.3, is  about  30  cylinders  across.  These  numbers  are
reasonable, so we take $f_A \approx  0.45$ as the lower bound on $f_A$
for  which the  theory is  valid.  For  $f_A <  0.452$ self-consistent
field-theory  predicts  that the  gyroid  phase  is  stable along  the
lamellar/cylinder  phase boundary.  Furthermore,  the weak-segregation
approximation will become increasingly inaccurate as $f_A$ decreases.

As the  mean-field critical point is  approached the trend  is for the
nucleation  barrier  to  decrease,  the  droplet  interface  width  to
increase, and  the droplet size to increase.   Thus the slowly-varying
amplitude  approximation  should   be  increasingly  accurate  as  the
critical point is approached.  We also have $l_{cx} > w_x$ and $l_{cz}
> w_z$ in this limit, which is a necessary condition for the classical
nucleation  and Wulff  approaches to  be  valid.  It  is possible  for
$l_{cz}   <  w_x$   for  some   super-heatings,  as   seen   in  Fig.\
\ref{FIG:LENGTHS}b, possibly indicating  a breakdown of this approach,
however one can go to smaller super-heatings where $l_{cz} > w_x$.  As
an example, for  $f_A = 0.49$ and  $\chi N = 10.5275$ with  $N = 1000$
and $\rho_0  = b^{-3}$, the nucleation  barrier is 28 $k_B  T$ and the
critical droplet, which has an aspect  ratio of about 10, is about 420
cylinders  across.  For these  values of  $f_A$ and  $\chi N$  we find
$l_{cz}  \approx 3.7$  $w_x$.   The interfacial  width for  interfaces
parallel to  the cylinders is  on the order  of 5--6 $D_c$,  while for
perpendicular interfaces the width is  on the order of 0.9 $D_c$. Near
the mean-field critical point  fluctuations will be important and will
renormalize  the basic  model,  Eq.\ (\ref{EQ:LB0}),  as discussed  in
\cite{BRAZOVSKII75,FREDRICKSON87,HOHENBERG95}.   Mean-field  theory is
recovered  in  the  limit   $N  \rightarrow  \infty$.   The  technique
discussed here can be applied  to study nucleation in the renormalized
model, in  which the lamellar/cylinder phase boundary  terminates at a
lamellar/cylinder/disorder  triple point  for $f_A  < 0.5$,  where the
nucleation barrier is  expected to remain finite \cite{FREDRICKSON87}.
In addition, we  expect that when the nucleation  barrier becomes less
than about  10 $k_B T$, either  near the mean-field  critical point or
the spinodal  curve, the  distinction between nucleation  and spinodal
decomposition will be lost and our approach will require modification.

We are not  aware of any experimental investigations  of the shape and
size  of nuclei in  transitions between  the lamellar  and cylindrical
phases in diblock  copolymer melts.  However, it is  worth noting some
experimental observations  on related  systems.  Koizumi {\em  et al.}
used   transmission  electron   microscopy   to  study   a  blend   of
poly(styrene-{\em   block}-isoprene)    and   homopolystyrene,   which
macrophase-separated to  form diblock-rich, lens-shaped  droplets in a
homopolymer-rich matrix  \cite{KOIZUMI94}.  Microphase separation into
a hexagonal  lattice of cylinders  occurred in the droplets,  with the
cylinder  axis   aligned  along  the  short  axis   of  the  lens-like
macrodomain, which had an aspect ratio of about 3.5.  Perpendicular to
the cylinder  orientation, the  droplet had an  approximately circular
cross-section  of  about  20   cylinders  in  diameter.   While  their
observations closely resemble our  results for the critical droplet at
$f_A =  0.45$, their  system is quite  different from  ours.  Firstly,
they   are    looking   at   macrophase    separation,   rather   than
nucleation. Secondly, in their  droplet the cylinders terminate at the
interface  with  the disordered,  homopolymer-rich  phase, while  ours
merge  continuously  into  a  lamellar  structure.   Thus,  while  the
relationship, if any, between these experiments and our theory remains
to  be understood,  it is  intriguing that  droplets similar  to those
predicted here exist in nature.

Balsara and coworkers have examined the evolution of cylindrical order
in  poly(styrene-{\em  block}-isoprene) following  a  quench from  the
disordered  state  \cite{DAI96,BALSARA98a,  BALSARA98b}.   From  their
depolarized light-scattering data they  inferred that their nuclei had
an aspect ratio of about 4, and that nucleation was occurring with the
cylinders oriented  along the  long axis of  the droplet.   While they
observe  an anisotropic  droplet shape,  their cylinders  are oriented
oppositely  to  those  in  the   present  theory,  and  those  in  the
experiments  of Koizumi  {\em et  al.}  \cite{KOIZUMI94}.   Of course,
since the  situations studied  are all different,  it is  difficult to
draw  any conclusions by  comparison.  However,  we have  repeated the
analysis  described  here  for  a  droplet  of  cylinder  phase  in  a
metastable disordered background  and our preliminary results indicate
that  the  cylinders   still  align  along  the  short   axis  of  the
droplet. Again, the interfacial  free-energy is lower when the droplet
interface lies  perpendicular to the cylinder  axis.  This discrepancy
with  the  conclusions  of  Refs.\  \cite{DAI96,BALSARA98a,BALSARA98b}
needs to be better understood.

Nonomura  and Ohta recently  performed two-dimensional  simulations of
the  nucleation and  growth  of a  droplet  of lamellar  phase from  a
metastable  cylinder phase  \cite{NONOMURA01}.  Although  this  is the
reverse  of the  transition  considered here,  comparison is  possible
since the  present theory for the interfacial  profile and free-energy
does not refer to which phase  is on the inside of the droplet.  Their
simulations, performed in what here is the $x$--$y$ plane, observe the
epitaxy we assume. Their droplet interface is relatively sharp, with a
width on the order of 2--3 cylinder spacings.  Their data suggest that
the droplet has a hexagonal  shape in the the $x$--$y$ plane, although
their larger  droplets, on  the order of  20 cylinders  across, appear
more circular. In  the absence of a precise  definition of the droplet
surface in the simulations, these observations about the droplet shape
can  only be  considered  qualitative. It  appears  that the  critical
droplet size in  the simulations is less than  1--2 cylinder spacings,
since droplets of  all sizes grow.  This small  critical droplet size,
less  than that  found  here,  may occur  because  of different  model
parameters  used  in  the  simulations,  or because  the  droplet  was
nucleated  heterogeneously  by  creating  a dislocation  pair  in  the
cylindrical order, thereby lowering the barrier to nucleation.

We have demonstrated the  crucial connection between anisotropy in the
interfacial free-energy and droplet shape, and Fig.\ \ref{FIG:BARRIER}
demonstrates the importance of using the proper critical droplet shape
to accurately  calculate the nucleation barrier. Thus  it is important
to consider  the robustness  of our results  for the  critical droplet
shape and  nucleation barrier.   We consider the  class of  models for
diblock  copolymers where  the free-energy  is written  in terms  of a
Landau expansion in the  order-parameter, and whose gradient structure
occurs at quadratic order in the order-parameter.  This class includes
the Landau-Brazovskii  model, the Leibler  model with local  cubic and
quartic  coefficients \cite{LEIBLER80},  and  the Ohta-Kawasaki  model
\cite{OHTA86}.   If  the   single-mode  and  slowly-varying  amplitude
approximations are applied  to this class of models  one can show that
the  resulting  gradient  structure  will  reduce to that of  Eq.\
(\ref{EQ:LBSINGLE}).   Since   the  critical  droplet   shape  depends
crucially on the gradient structure  of the theory, this suggests, for
weak segregation, that the lens-like critical droplet shape calculated
here  is  robust. At  higher  segregation  the  droplet shape  may  be
modified from that calculated here.

The need  to use a variational  ansatz to obtain the  free-energy of a
planar  interface can be  eliminated by  solving the  amplitude model,
Eqs.\   (\ref{EQ:EL1})  and   (\ref{EQ:EL2}),  numerically   in  three
dimensions  for  a  droplet  of  cylinder phase  in  a  background  of
metastable  lamellar phase.   Near the  phase boundary,  we  expect no
qualitative change in either the droplet shape or the behaviour of the
nucleation barrier from that  reported here.  The overall magnitude of
the nucleation  barrier will, however,  be somewhat reduced  since the
elimination of the variational ansatz  will lead to a reduction in the
calculated interfacial free-energy. This  approach will also enable us
to go beyond the constraints imposed by the Wulff construction and the
classical  nucleation theory. It  will allow  an investigation  of the
entire  metastable lamellar  region  from the  phase  boundary to  the
spinodal.   The  nucleation  barrier  should  approach  zero  and  the
critical   droplet   dimensions  should   diverge   at  the   spinodal
\cite{CAHN59}.   As  the  spinodal   is  approached  and  the  droplet
interface  becomes  more diffuse  the  connection between  interfacial
free-energy  and droplet shape  becomes less  clear, thus  the droplet
shape may be  modified near the spinodal.  It may  be possible to make
contact  with  other  theories  that  address the  spinodal  limit  of
order-order transitions \cite{LARADJI97,  SHI99, QI97, QI98, MATSEN98,
MATSEN01, QI96}.

In  addition  to  exploring  the  limits of  the  theory  through  the
extensions  discussed  above,  we  plan  to  apply  the  formalism  to
nucleation at the sphere/cylinder  transition. The role of anisotropic
fluctuations during this transition has been studied experimentally in
Ref.\ \cite{RYU98}.  This transition has been studied theoretically in
Refs.\  \cite{QI96,LARADJI97,MATSEN01}  assuming uniform  intermediate
states, but the nucleation of  compact droplets has not been examined.
Finally,  we would  like  to  develop a  deeper  understanding of  the
orientation dependence of the  interfacial free-energy in terms of the
structure of the lamellar/cylinder  interface and the conformations of
block chains at these interfaces.
%
%
\section{Conclusion}

We have examined the nucleation  of a droplet of stable cylinder phase
from a  metastable lamellar phase using  the single-mode approximation
to  the  Brazovskii  model  for  diblock  copolymer  microphases.   By
employing a variational ansatz for the droplet interfacial profile, we
find  an analytic  expression for  the interfacial  free-energy  of an
interface of arbitrary orientation between cylinders and lamellae. The
interfacial free-energy is anisotropic, and is lower when the cylinder
axis is  perpendicular to  the interface than  when the  cylinders lie
along  the  interface.  Furthermore,  the  interfacial free-energy  is
slightly  lower  when  the  lamellae  are parallel  to  the  interface
compared to  the perpendicular alignment.  The  droplet shape computed
{\em via}  the Wulff construction is lens-like,  being flattened along
the  axis of  the  cylinders.   As the  mean-field  critical point  is
approached  along  the lamellar/cylinder  phase  boundary the  droplet
becomes  more  flattened  along  the  cylinder axis.   We  apply  this
information  to compute  the  size  of the  critical  droplet and  the
nucleation barrier within classical nucleation theory.  We are able to
make  specific  predictions for  these  quantities  by connecting  the
phenomenological   Brazovskii   model   to  the   many-chain   Edwards
Hamiltonian  for diblock  copolymers.  The general  trend  is for  the
nucleation  barrier  to decrease  and  the  critical  droplet size  to
increase  as  the  mean-field   critical  point  is  approached.   The
nucleation barrier is significantly  reduced when the critical droplet
shape is  anisotropic instead of spherical,  indicating the importance
of using  the proper droplet shape.   The theory should  be valid near
the lamellar/cylinder phase boundary from the lamellar/gyroid/cylinder
triple point  at $f_A  \approx 0.45$ to  near the  mean-field critical
point. In this regime, droplets  of size 30--400 cylinders across with
aspect ratios of  4--10 and nucleation barriers of  30--40 $k_B T$ are
typically found.  Close to  the mean-field critical point fluctuations
may  modify   this  mean-field   picture.   Due  to   the  variational
approximation, the computed interfacial free-energies are upper bounds
on the true free-energies, implying that the computed critical droplet
dimensions are also upper bounds on the true dimensions.

Our  theory  makes  specific predictions  about  the
nucleation barrier and the size  and shape of the critical droplet for
nucleation  of the  cylindrical phase  from metastable  lamellae in
diblock copolymer melts.  We systematically
calculate  the  interfacial free-energy  for  interfaces of  arbitrary
orientation  between lamellae and  cylinders. This work is an 
important first step towards a more sophisticated  theory of
nucleation in this system.  The  size of the critical droplets we find
already  suggests that  a direct  numerical attack  on  the nucleation
problem, using Eq.\ (\ref{EQ:LB0}),  will be challenging. Although our
focus  has been  on nucleation  in diblock  copolymers,  this approach
should work for  any system in the Landau-Brazovskii  class, given the
appropriate model parameters.   The nucleation scenario described here
should be  observable experimentally, however  experiments which study
droplet  nucleation  along  the  lamellar/cylinder phase  boundary  in
diblock copolymer melts have yet to be performed.
%
%
\acknowledgments

The authors would like to thank Professor Martin Grant for helpful comments.
This work was supported by the Natural Sciences and Engineering Research 
Council of Canada, the Research Corporation, and an Ontario Premier's Research
Excellence Award. ZGW acknowledges support by 
the US National Science Foundation (DMR-9970589).

\end{document}